\documentstyle[aps,epsf,epsfig,prl,multicol]{revtex}
\begin{document}
\draft

\title{Avoiding Quantum Chaos in Quantum Computation}

\author{G.P. Berman$^{[a]}$,F. Borgonovi$^{[b,c]}$,
F.M. Izrailev$^{[d]}$, V.I. Tsifrinovich$^{[e]}$ }
\address{
$^{[a]}$Theoretical Division and CNLS, Los Alamos National
Laboratory, Los Alamos, New Mexico 87545\\ $^{[b]}$Dipartimento di
Matematica e Fisica, Universit\`a Cattolica, via Trieste 17, 25121
Brescia, Italy  \\ $^{[c]}$ I.N.F.M., Gruppo Collegato di Brescia
and I.N.F.N., Sezione di Pavia, Italy\\ $^{[d]}$Instituto de
Fisica, Universidad Autonoma de Puebla, Apdo. Postal J-48, Puebla
72570, Mexico\\ $^{[e]}$ IDS Department, Polytechnic University,
Six Metrotech Center, Brooklyn, NY 11201\\ }

\maketitle

\maketitle

\begin{abstract}
We study a one-dimensional chain of nuclear $1/2-$spins in an
external time-dependent magnetic field. This model is considered
as a possible candidate for experimental realization of quantum
computation. According to general theory of interacting particles,
one of the most dangerous effects is quantum chaos which can
destroy the stability of quantum operations. The standard
viewpoint is that the threshold for the onset of quantum chaos due
to an interaction between spins (qubits) strongly decreases with
an increase of the number of qubits. Contrary to this opinion, we
show that the presence of a non-homogeneous magnetic field can
strongly reduce quantum chaos effects. We give analytical
estimates which explain this effect, together with numerical data
supporting our analysis.
\end{abstract}
\date{30.11.2000}
\begin{multicols}{2}

Much attention is paid in recent years to the idea of quantum
computation (see, for example, \cite{S98,BDMT98,chuang} and
references therein). The burst of interest to this subject is
caused by the discovery of fast quantum algorithm for the
factorization of integers \cite{S94} demonstrating the
effectiveness of quantum computers in comparison to the classical
ones. Nowadays, there are different projects for the experimental
realization of quantum computers, based on interacting two-level
systems (qubits). One of the most important problems widely
discussed in the literature, is the problem of decoherence which
arises in many-qubit systems due to the influence of an
environment \cite{CLSZ95}. However, even in the absence of the
environment, the interaction between qubits may lead to the
``internal decoherence" related to the onset of quantum chaos
\cite{dima}.

The latter subject of quantum chaos in closed systems of
interacting particles has been developed recently in application
to nuclear, atomic and solid state physics (see, e.g.,
\cite{GMW99} and references therein).
When the (two-body) interaction between particles
exceeds the critical value,
fast transition to chaos occurs in the
Hilbert space of many-particle states \cite{AGKL97}. Different aspects of this
transition are now well understood, such as statistical
description of eigenstates and the onset of thermalization in
finite systems (see, e.g., \cite{I00} and references therein).

Direct application of the quantum chaos theory to a  simple model
of quantum computer \cite{dima} has shown that for a strong enough
interaction between qubits the onset of quantum chaos is
unavoidable. Although for $L=14-16$ qubits the critical value
$J_{cr}$ for quantum chaos threshold is quite large, with an
increase of $L$ it decreases as $J_{cr} \sim 1/L$. From the
viewpoint of the standard approach for closed systems of
interacting particles, the decrease of the chaos threshold with an
increase of qubits looks generic. However, in this Letter we
demonstrate that this conclusion is not universal and the quantum
chaos can be avoided, for example, with a proper choice of the external magnetic
field.

Our consideration is based on the one-dimensional model of $L$
nuclear $1/2-$spins subjected to the time-dependent magnetic field
of the following form
\cite{gena},
\begin{equation}
{\vec B}(t)=[b^p_\perp\cos(\nu_p t+\varphi_p),-b^p_\perp\sin(\nu_p t+\varphi_p), B^z],
\label{sp}
\end{equation}
where $B^z$ is a constant magnetic field
oriented in the positive $z$-direction, $b^p_\perp$ , $\nu_p$, and $\varphi_p$
are the amplitudes, frequencies and phases of a circular polarized
magnetic field which is given by the sum of $p=1,...,P$
rectangular pulses of the length $t_{p+1}-t_p$ rotating in the $(x,y)-$ plane, and providing a quantum computer protocol. The quantum
Hamiltonian of this system has the form,
\begin{equation}
\begin{array}{ll}
{\cal H}= -\sum\limits^{L-1}_{k=0} (\omega_kI^z_k+2
\sum\limits_{n > k}J_{k,n} I^z_k I^z_n)-
\\ ~~~~~~~~~\\
{{1}\over{2}}\sum\limits_{p=1}^{P}\Theta_p(t)\Omega_p
\sum\limits_{k=0}^{L-1} \Bigg(e^{-i\nu_p
t-i\varphi_p}I^-_k+ e^{i\nu_p t+i\varphi_p}I^+_k\Bigg),
\label{ham00}
\end{array}
\end{equation}
where the ``pulse function" $\Theta_p(t)$ equals $1$ only during
the $p$-th pulse, for $t_p< t\le t_{p+1}$. The quantities
$J_{k,n}$ stand for the Ising interaction between two qubits ,
$\omega_k$ are the frequencies of spin's precession in the
$B^z-$magnetic field, $\Omega_p$ is the Rabi frequency of the
$p$-th pulse, $I_k^{x,y,z} = (1/2)
\sigma_k^{x,y,z}$, the latter being the Pauli matrices, and
$I_k^{\pm}=I^x_k \pm iI^y_k$.

For the $p$-th pulse, the Hamiltonian (\ref{ham00}) can be represented in the coordinate
system which rotates with the frequency $\nu_p$. Thus, for the $p$-th pulse, our model can be reduced to the {\it stationary} Hamiltonian,
\begin{equation}
\begin{array}{ll}
{\cal H}^{(p)}=-\sum\limits_{k=0}^{L-1} [(\omega_k-\nu_p)I^z_k+
\Omega_p(\cos\varphi_p I^x_k-\sin\varphi_p I^y_k)+
\\ ~~~~~~~~~\\
2 \ \sum\limits_{n>k}^{}J_{k,n}I^z_k I^z_n],
\label{ham}
\end{array}
\end{equation}
which describes the evolution for $t_p<t\le t_{p+1}$.

The physically important regime of quantum computation is a {\it
selective excitation} which corresponds to the range of
parameters: $\Omega_p\ll J_{k,n}\ll\delta\omega_k\ll\omega_k$, where $\delta\omega_k=|\omega_{k+1}-\omega_k|$
\cite{large}. However, in this Letter, we consider a regime of
{\it non-selective excitation} which is defined by the conditions,
$\Omega_p\gg \delta\omega_k \gg J$, see details in \cite{gena}.
This inequality provides the simplest way to prepare a homogeneous
superposition of $2^L$ states needed for implementation of both
Shor and Grover algorithms. In what follows we assume, for
simplicity, that the amplitudes of magnetic pulses, and their
frequencies and phases are equal, $\Omega_p=\Omega$, $\nu_p=\nu$,
and $\varphi_p=\pi/2$. Our main interest is in the nearest
neighbor interaction ({\it N-interaction}) between qubits for two
different cases, the {\it dynamical} one when all coupling
elements are the same, $J_{k,n}=J\ \delta_{n,k+1}$, and the case
when all values $J_{k,k+1}$ are random ({\it random model}).
However, we will also discuss another case when all qubits
interact to each other ({\it A-interaction}) with random
$J_{k,n}$. In contrast to previously discussed model \cite{dima}
with homogeneous magnetic field, below we consider the magnetic
field with constant gradient along the z-axes. Therefore, we
assume that the spin frequencies $\omega_k$ are slightly dependent
on $k$ in the following way, $\delta\omega_k  \ll \omega_k$. For the dynamical $N-$interaction and
$\nu=\omega_0,\,\omega_k=\omega_0 +ak$ ($a>0$), the Hamiltonian takes the
form,
\begin{equation}
H=\sum_{k=0}^{L-1}
\Big [-\delta_kI^z_k+
\Omega I^y_k\Big] -2J\sum_{k=0}^{L-2} I^z_k I^z_{k+1}.
\label{ham0}
\end{equation}
where $\delta_k=ak$. In the $z$-representation the Hamiltonian
matrix of size $N = 2^L$ is diagonal for $\Omega=0$. For
$\Omega\not=0$, $H_{k,k+n}=i\Omega/2$  and $H_{nk}=H^*_{kn}$. The
matrix is very sparse, and it has quite specific structure if the
basis is reordered according to an increase of the number $s$
written in the binary representation, $s=i_1,i_2,...,i_{L}$ (with
$i_s=0$ or $1$, depending on whether a single-particle state of
$i-$th qubits is the ground state or the excited state).

The structure of eigenstates in this basis has been numerically
studied in detail for $L=8,10,12$ in dependence on the strength of
interaction, $J$, and different values of $\Omega$. For $\Omega
\gg J $ the spectrum consists of $L+1$ narrow bands with large
gaps of size $\approx \Omega$ between the bands. Since the most
interesting energy region for preparation of the homogeneous wave
function is in the middle of the energy (quasi-energy) spectrum,
we consider below only the central band and the corresponding
eigenstates for $L$ even.

\begin{figure}
\epsfxsize 7cm
\epsfbox{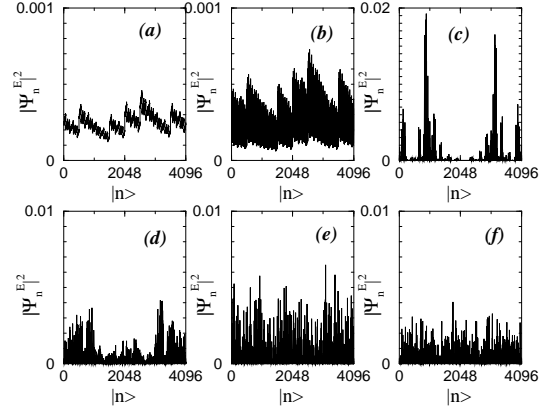}
\narrowtext
\caption{Typical structure of eigenfunctions for
$J=0,\, 2\cdot 10^{-4},\, 0.1,\,1,\, 10,\, 100 $ denoted by
(a,b,c,d,e,f) respectively. Eigenstates are taken from the central
energy band for $L=12,\, \Omega=100,\, a=1$.}
\label{eig}
\end{figure}

In the absence of interaction between qubits, $J=0$, all
eigenstates in the given basis are fully extended with the value
of components $\psi_n$ close to $1/\sqrt N$. Typical structure of
eigenstates in the central band is shown in Fig. \ref{eig} for
different values of $J$. One can see that with an increase of
$J-$interaction, the probabilities, $w_n=|\psi_n|^2$, deviate from
the unperturbed value, $w_n=1/N$ , thus resulting in quantum
computation errors. The data demonstrate the transition from
regular states for a weak interaction, $J \leq 0.1$, to {\it
chaotic} ones for $J \approx 100$.

Global properties of eigenstates can be characterized by the
number, $N_{pc}$, of principal components, determined through the
{\it inverse participation ratio}, $N_{pc}(E) =
\left[\sum_n |\psi_n(E)|^4\right ]^{-1}$, where $\psi_n(E)$
is the $n$-th component of a specific eigenfunction. For zero
interaction ($J=0$), the number $N_{pc}$ is equal to $N$, and it
decreases with an increase of interaction, thus, giving the
measure of the destruction of unperturbed ($J=0$) eigenstates.
Note that for completely chaotic eigenstates with gaussian
fluctuations for $\psi_n$, one has $N_{pc}=N/3$, \cite{GMW99}.

\begin{figure}
\epsfxsize 7cm
\epsfbox{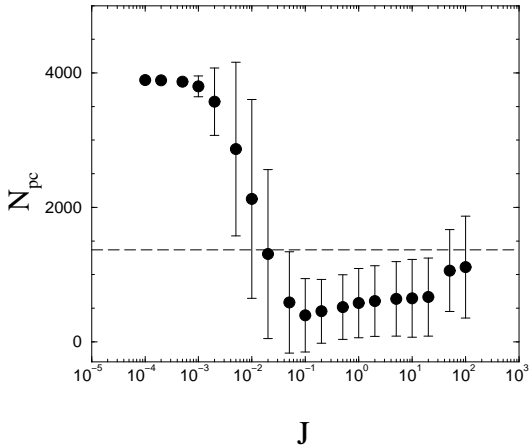}
\narrowtext
\caption{
The average number of principal components $N_{pc}$  as a function of
$J$ for the parameters of Fig. \ref{eig}; the horizontal line
corresponds to $N/3$. The average is taken over the eigenfunctions
from the central band only.}
\label{ll}
\end{figure}

Numerical data for the averaged $N_{pc}$ in dependence on the
interaction $J$  define different regions, see Fig.
\ref{ll}. The first region on the left of this figure corresponds to
completely ``delocalized" eigenstates with $\psi_n \approx
1/\sqrt{N}$. Here the energy spectrum is characterized by many
close quasi-degenerate levels. In the second region where
$N_{pc}\ll N$, all eigenstates are strongly influenced by the
inter-qubit interaction. In this region which we denote as the
region of {\it weak chaos}, the level spacing distribution $P(s)$
is quite close to the Poisson. From the data, the transition to
the region of weak chaos occurs for $J = J_{cr}\approx 0.05$. One
should stress that from the practical point of view the {\it weak
chaos} should be avoided in the quantum computation because of
large deviations of eigenstates from the unperturbed ones, see
Fig. 1c-d.

Second transition to a {\it strong quantum chaos} occurs for $J
\geq 20$. By the latter term we denote the situation when the
level spacing distribution has the Wigner-Dyson form and
fluctuations of components $\psi_n$ are close to the gaussian ones
with $N_{pc} \approx N/3$. Analysis shows that this transition
corresponds to the overlapping of the central energy band to the
nearest ones. As one can see, strong quantum chaos emerges for
extremely strong interaction. Quite unexpectedly, the results for the {\it
random} model (with random distribution of the interaction
strength in the interval $[-J, J]$) turn out to be the
same. This means that, both for the dynamical and random
$N$-interactions our system is close to an integrable one for well
separated energy bands.

The above {\it numerical} analysis for $L=12$ shows that weak
quantum chaos can significantly influence the structure of
eigenstates, but the regime of strong chaos is not achievable in
quantum computation. However, according to a common believe, the
threshold for {\it both} weak and strong chaos is expected to
decrease with an increase of $L$, thus, leading to the destruction
of unperturbed states even in the presence of a very weak
interaction $J$.

We show now {\it analytically}, that contrary to the standard
point of view, in this model the chaos border $J_{cr}$ {\it is
independent} on the number of qubits. In order to explain this
unexpected phenomena, it is convenient to represent the
Hamiltonian (\ref{ham0}) in the basis in which it is diagonal for
non-interacting ($J=0$) qubits,
\begin{equation}
H= H_0 + JV_0.
\label{H0V}
\end{equation}
Here the ``effective field'' Hamiltonian, $H_0$, is determined by
the sum of $L$ individual Hamiltonians $H_k$,
\begin{equation}
H_0 = \sum\limits_{k=0}^{L-1} H_k = \sum\limits_{k=0}^{L-1}
\sqrt{ (\omega_k-\nu)^2 +
\Omega^2 } \,\, I_k^z,
\label{H0}
\end{equation}
and the interaction $V_0=V_{diag} + V_{band} + V_{off}$ between
new ``quasi-particles" has the form,
\begin{equation}
\begin{array}{ll}
V_{diag} = -2 \sum\limits_{k} b_k b_{k+1} I^z_k I^z_{k+1}
\,;\,\,\,
\\ ~~~~~~~~ \\
V_{band} = -2 \sum\limits_{k} a_k a_{k+1} I^y_k I^y_{k+1}
\,;\,\,\,
\\ ~~~~~~~~ \\
V_{off}  = 2 \sum\limits_{k}
\left(  a_k b_{k+1} I^y_k I^z_{k+1} +
a_{k+1} b_{k} I^z_k I^y_{k+1} \right).
\end{array}
\label{Vvd}
\end{equation}
Three terms in the expression for $V_0$ stand for the
diagonal, inside-band and between-bands interaction respectively,
and
\begin{equation}
b_k =
\frac{\nu-\omega_k}{\sqrt{ (\omega_k-\nu)^2 + \Omega^2 }}; \,\,\,\,\,
a_k = \frac{\Omega} {\sqrt{ (\omega_k-\nu)^2 + \Omega^2 }}.
\label{ak}
\end{equation}

Under the condition, $|\omega _k-\nu| \ll \Omega$, we have the
following expression for single-particle ``quasi-energies''
corresponding to the Hamiltonian $H_0$,
\begin{equation}
\label{eps}\epsilon _k=\pm\frac{1}{2}\sqrt{(\omega _k-\nu )^2+\Omega ^2}
=\pm \frac 12\left( \Omega
\;+\frac{a^2 k^2}{2\Omega }\right).
\label{plusminus}
\end{equation}
Note, that each ``quasi-particle" can have $2L$ different
``quasi-energies" $\epsilon_k$ in contrast to actual particles
(qubits) which have only two values of energy. The above
expression allows us easily to construct many-particle unperturbed
quasi-energies, $E_c=\sum_{k=0}^{L-1}\epsilon _k$, inside the
central band. Indeed, since we have assumed $\Omega \gg |\omega
_k-\nu |$ all many particles levels should have, for $L$ even,
$L/2$ positive $\epsilon _k$ and $L/2$ negative ones. As a result,
the total number $N_{cb}$ of many-body states in the central band
is given by $N_{cb}=L!({\frac L2!\frac L2!})^{-1}$ and the size
can be estimate as twice the maximum energy, $(\Delta E)_{cb} = 2
E_c^{(max)}= a^2 L^2(L-1)/8\Omega$.

Now, we can estimate, for the $N-$interaction, the mean level
spacing $\delta E$ between those many-body states which are
directly coupled by the interaction, see Eq.(\ref{Vvd}). The
value $\delta E$ can be estimated as the ratio $\delta E= (\Delta
E)_f /M_f$ where $M_f\approx L/2$ is the number of many-body
states coupled by $V_{band}$ (it is, in fact, the mean number per
line of non-zero off-diagonal elements in the total Hamiltonian
(\ref{H0V})). One should stress that the energy range $(\Delta
E)_f$ within which these states are coupled, is less than the
total energy width $(\Delta E)_{cb}$ of the central band. The
value of $(\Delta E)_f$ can be estimated as the maximal difference
between energies $E_c^{(2)}=\sum_k^{(2)}
\epsilon_k $ and $E_c^{(1)}=\sum_k^{(1)}
\epsilon_k$ of two many-body states $|1\rangle$ and $|2\rangle$ of $H_0$, that
have the coupling $\langle 1|V_{band}|2\rangle$ different from zero. From the
expression (\ref{plusminus}) one finds, $(\Delta E)_f =
2a^2L/\Omega$. As a result, for $L \gg 1$ we have,
\begin{equation}
\delta E = \frac {(\Delta E)_f}{M_f} \approx
\frac{2a^2}{\Omega}
\label{deltae}
\end{equation}

This mean spacing, $\delta E$, should be compared with the typical
value of perturbation $V=JV_0$ \cite{I00}. The latter can be found
from $V_{band}$ as $V \approx J/2$. Therefore, we finally obtain,
\begin{equation}
J_{cr} \approx \frac{4a^2}{\Omega}.
\label{J1}
\end{equation}

Remarkably, the threshold to chaos, $J_{cr}$, does not depend on the number $L$ of qubits. The
origin of this phenomenon is that the width $(\Delta E)_f$ and
$M_f$ both increase lineraly in $L$. This fact is entirely due to
the quadratic growth of the energy shift in Eq. (\ref{eps}) for
new single-particles states, in dependence on $k$ . Physically,
this effect is related to the constant gradient of the magnetic
field in the original Hamiltonian.

Numerical data for the number $N_{pc}$ of principal components of
eigenstates inside the central energy band in the new basis (where
$H_0$ is diagonal for $J=0$, see Eq. (\ref{H0V})) are given in
Fig. 3 in dependence on $J/J_{cr}$. It is clearly seen that below
the border, $J < J_{cr}$, there is a scaling dependence of
$N_{pc}$ on $L$ and $\Omega$ which confirms our estimate
(\ref{J1}). On the other side, for $J > J_{cr}$, the value of
$N_{pc}$ saturates to its maximal value $N_{cb}/3$ in
correspondence with random matrix predictions \cite{GMW99}. We
notice that the transition to extended states in the Hamiltonian
(\ref{H0V}), which occurs for $J > J_{cr}$, corresponds to the
transition to the {\it weak chaos} given in Fig.2.

In the case of {\it random} model, the data turns out to be
similar, thus, confirming our conclusion that for the
$N$-interaction the model is close to an integrable one,
independently on whether the interaction is dynamical or random.
One should note that, indeed, if one neglects the interband
interaction $V_{off}$, the Hamiltonian (\ref{H0V}) is integrable,
as shown in \cite{Young}. Thus, our analysis shows that the border
$J_{cr}$ does not decrease with an increase of number of qubits.

\begin{figure}
\epsfxsize 7cm
\epsfbox{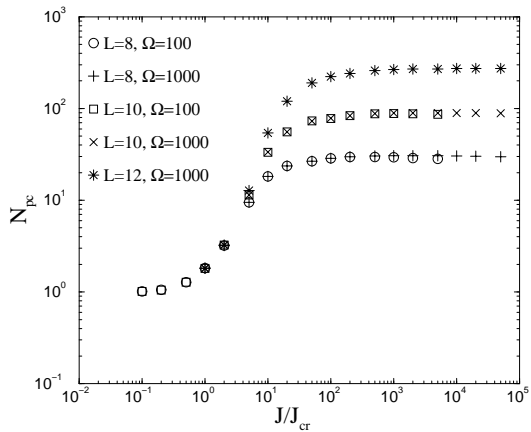}
\narrowtext
\caption{
The average number of principal components for the dynamical
$N$-interaction as a function of $J/J_{cr}$, in the central band
in the $H_0-$basis, for $L=8,\,10,\,12$ and different values of
$\Omega$.}
\label{dyn}
\end{figure}

Now we have to take into account that with an increase of $L$, the
width $(\Delta E)_{cb}$ of the central band can exceed the
distance $\Omega$ between the nearest ones, which leads to strong
quantum chaos. The estimate for the critical value, $L_{cr}$,
which corresponds to the band overlapping in absense of the
$J-$interaction reads as $L_{\max } \sim (\Omega/a) ^{2/3} \gg 1$.

For the model with random interaction between all qubits
($A$-interaction),  strong quantum chaos occurs {\it inside} the
energy bands  manifested by a  Wigner-Dyson distribution for
$P(s)$ (more details will be given elsewhere). In this case the
estimate for the energy width $(\Delta E)_f$, within which
many-body states are coupled by the two-body interaction, is
$(\Delta E)_f = L^2/2\Omega$, and $M_f =a^2 L^2/4$. Therefore, we
get $\delta E = (\Delta E)_f / M_f \approx 2a^2/\Omega$ which is
the same as for the $N-$interaction. This is an unexpected result
since generically, the chaos border for $A-$interaction is much
less than for $N-$interaction ($J_{cr}\propto 1/L^2$, see, e.g.
\cite{dima0}). Numerical data for the $A-$interaction in
(\ref{ham0}) confirm the above result (the corresponding figure is
very similar to Fig. 3).

In conclusion, we have shown that, in contrast to general believe,
the chaos border in the model of $L$ interacting qubits does not
decrease with an increase of $L$, in the presence of a strong
magnetic field with constant gradient in the $z$-direction.
The quantum chaos which emerges for a very strong
interaction between qubits is irrelevant to quantum computation.
The mechanism of strong chaos is shown due to the band overlapping
only, and can be avoided even for a very large number of qubits.
Moreover, our analysis shows that for an inhomogeneous gradient of
magnetic field (for example, when $\omega_k
\propto k^4$), one can expect that the border for chaos
{\it increases} with an increase of $L$. The region of parameters for quantum
computation with {\it selective excitation} requires additional
analysis.

The work  of GPB and VIT was supported by the Department of Energy
(DOE) under the contract W-7405-ENG-36, by the National Security
Agency (NSA) and Advanced Research and Development Activity
(ARDA). FMI acknowledges the support  by CONACyT (Mexico) Grant
No. 34668-E.

\end{multicols}

\end{document}